\documentclass[twocolumn,showpacs,preprintnumbers,amsmath,amssymb]{revtex4}

\usepackage{graphicx}
\usepackage{dcolumn}
\usepackage{bm}

\begin{document}

\preprint{APS/123-QED}

\title{Origin of the material dependence of $T_c$ in the 
single-layered cuprates}

\author{Hirofumi Sakakibara$^1$}
\author{Hidetomo Usui$^2$}
\author{Kazuhiko Kuroki$^{1,5}$}
\author{Ryotaro Arita$^{3,5,6}$}
\author{Hideo Aoki$^{4,5}$}

\affiliation{$\rm ^1$Department of Engineering Science, The University of Electro-Communications, Chofu, Tokyo 182-8585, Japan}
\affiliation{$\rm ^2$Department of Applied Physics and Chemistry, The University of Electro-Communications, Chofu, Tokyo 182-8585, Japan}
\affiliation{$\rm ^3$Department of Applied Physics, The University of Tokyo, Hongo, Tokyo 113-8656, Japan}
\affiliation{$\rm ^4$Department of Physics, The University of Tokyo, Hongo, Tokyo 113-0033, Japan}
\affiliation{$^5$ JST, TRIP, Sanbancho, Chiyoda, Tokyo 102-0075, Japan}
\affiliation{$^6$ JST, PRESTO, Kawauchi, Saitama 332-0012, Japan}

\date{\today}

\begin{abstract}
In order to understand the material dependence of $T_c$ 
within the single-layered cuprates, we study a two-orbital model that 
considers both $d_{x^2-y^2}$ and $d_{z^2}$ orbitals.  
We reveal that a hybridization of $d_{z^2}$ on the Fermi surface substantially 
affects $T_c$ in the cuprates, where 
the energy difference $\Delta E$ between 
the $d_{x^2-y2}$ and $d_{z^2}$ orbitals is identified to be 
the key parameter that governs 
both the hybridization and the shape of the Fermi surface.    
A smaller $\Delta E$ tends to suppress $T_c$ through a 
larger hybridization, whose effect supersedes the effect 
of diamond-shaped (better-nested) Fermi surface.  
The mechanism of the suppression of $d$-wave superconductivity due to 
$d_{z^2}$ orbital mixture is clarified from the viewpoint of the ingredients 
involved in the Eliashberg equation, 
i.e., the Green's functions and the form of the pairing 
interaction described in the orbital representation.
The conclusion remains qualitatively the same if we 
take a three-orbital model that 
incorporates Cu 4s orbital explicitly, where the 4s orbital is shown to have 
an important effect of making the Fermi surface rounded.  
We have then identified the origin of the material and 
 lattice-structure dependence of $\Delta E$, 
which is shown to be 
determined by the energy difference $\Delta E_d$ between the two Cu3d orbitals (primarily governed by the apical oxygen height), 
and the energy difference $\Delta E_p$ 
between the in-plane and apical oxygens (primarily governed by the interlayer separation $d$).  

\end{abstract}

\pacs{74.20.-z, 74.62.Bf, 74.72.-h}
\maketitle

\section{INTRODUCTION}

Despite the fact that the history of the 
high-$T_c$ cuprates exceeds two decades, 
there remain a number of fundamental 
questions which are yet to be resolved.
Among them is the 
significant variation of $T_c$ among various materials 
within the cuprate family.  
It is well known that 
$T_c$  varies strongly with the number of CuO$_2$ layers, 
but an even more 
basic problem is the $T_c$ variation 
within the single-layered materials.  This is highlighted 
by La$_{2-x}$(Sr/Ba)$_x$CuO$_4$ with a $T_c \simeq 40$ K 
versus HgBa$_2$CuO$_{4+\delta}$ with a $T_c \simeq 90$K, 
with a more than factor of two difference despite similar 
crystal structures between them.

Empirically, it has been recognized that the materials 
with $T_c\sim 100$K tend to have ``round'' Fermi surfaces,
while the Fermi surface of the 
La system is closer to a diamond shape, and this has posed a 
long-standing, big puzzle, since, the latter would imply a relatively 
better nesting\cite{Pavarini,Tanaka}.
The materials with rounded Fermi surfaces conventionally have been 
analyzed with a single-band model with large second ($t_2(>0)$) and 
third ($t_3(<0)$) neighbor hopping integrals,  
while the ``low-$T_c$'' La system has been considered to have 
smaller $t_2, t_3$. 
This, however, has brought about a contradiction between 
theories and experiments. 
Namely, while some phenomenological\cite{Moriya} 
and $t$-$J$ model\cite{Shih,Prelovsek} studies 
give a tendency consistent with the experiments, 
a number of many-body 
approaches for the {\it Hubbard-type} models with 
realistic values of on-site interaction 
$U$ show suppression of superconductivity for 
large $t_2>0$ and/or $t_3<0$, 
as we shall indeed confirm below\cite{Scalapino}.

To resolve this discrepancy, 
we have introduced  in ref.\cite{SakakibaraPRL} 
a two-orbital model that explicitly incorporates the $d_{z^2}$ orbital 
as well, while the usual wisdom was that the $d_{x^2-y^2}$ orbital 
suffices.  The former component 
has in fact 
a significant contribution to the Fermi surface in the 
La system. 
We have  shown that 
the key parameter that determines $T_c$ is the 
hybridization of the two-orbitals, 
which is in turn governed by the 
level offset $\Delta E$ between 
the $d_{x^2-y^2}$ and $d_{z^2}$ Wannier orbitals.  
Namely,  the weaker the $d_{z^2}$ contribution to the 
Fermi surface, the better for $d$-wave superconductivity, 
where 
a weaker contribution of the $d_{z^2}$ results in a rounded 
Fermi surface (which in itself is not desirable for superconductivity), 
but it is the ``single-orbital nature'' that favors a higher $T_c$ 
superseding the effect of the Fermi surface shape.  
Recently, there have also been some other theoretical studies 
regarding the role of the $d_{z^2}$ orbital played in the cuprates\cite{Kotliar,Millis,Fulde,Honerkamp}.

Purpose of the present paper is two-fold:  
By elaborating the two-orbital model, 
we investigate (i) why the $d_{z^2}$ hybridization 
on the Fermi surface
suppresses the superconductivity, 
and (ii) what are the key components 
that determine the material dependence 
in the level offset between $d_{x^2-y^2}$ and $d_{z^2}$. 
In examining point (ii), in addition to La$_2$CuO$_4$ and 
HgBa$_2$CuO$_{4+\delta}$ considered in ref.\cite{SakakibaraPRL}, 
we also construct effective models of 
the single-layered cuprates Bi$_2$Sr$_2$CuO$_6$ and Tl$_2$Ba$_2$CuO$_6$ to reveal how these materials can be classified in terms of the 
correlation between the lattice structure parameters and the 
level offsets of various orbits.

\section{Construction of the two-orbital model}
\subsection{Band calculation}
\label{IIA}

\begin{figure}[!b]
\includegraphics[width=7cm]{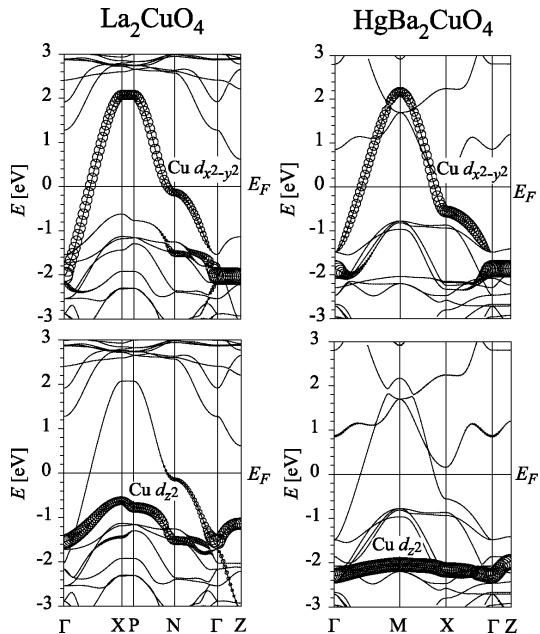}
\caption{First-principles band structures of  La$_2$CuO$_4$ (left) and 
HgBa$_2$CuO$_4$ (right). The top (bottom) panels 
depict the strength of the $d_{x^2-y^2}$ 
($d_{z^2}$) characters with the radius of the circles.  
}
\label{fig1}
\end{figure}

\begin{figure}[!b]
\includegraphics[width=7cm]{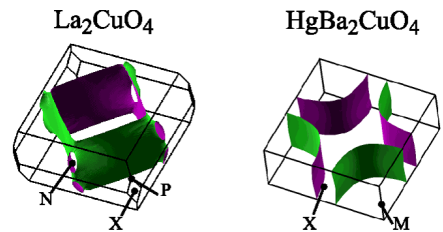}
\caption{(Color online) The Fermi surface of the  La$_2$CuO$_4$ (left) and 
HgBa$_2$CuO$_4$ (right) with 0.15 holes/Cu atom.  }
\label{fig2}
\end{figure}

Let us start with the first-principles band calculation\cite{wien2k} of 
La$_2$CuO$_4$ and HgBa$_2$CuO$_4$, whose 
band structures are displayed in Fig.\ref{fig1}. 
The lattice parameters adopted here are experimentally 
determined ones for the doped materials\cite{La-st,Hg-st}.
In both cases, there is only one band intersecting the  Fermi level. 
Therefore, the $d_{x^2-y^2}$  single-orbital 
 Hubbard model, 
or the Cu-$d_{x^2-y^2}$ $+$ O-$_{p\sigma}$ three-orbital model 
(whose antibonding band crosses the Fermi level)
has been adopted in conventional theoretical studies. 
A large difference between the two materials in the 
shape of the Fermi surface is confirmed in Fig.\ref{fig2}.
As mentioned in the Introduction, the materials with a rounded Fermi surface 
have been modeled by a single-orbital model 
with large second [$t_2(>0)$] and third [$t_3(<0)$] 
neighbor hopping integrals\cite{Pavarini}.
It has been noticed that when the fluctuation exchange  approximation 
(FLEX)\cite{Bickers,Dahm} is applied to this model, 
a rounder Fermi surface coming from larger second and third neighbor 
hoppings results in a suppressed $T_c$, as we have shown in 
Fig.1 of ref.\onlinecite{SakakibaraPRL}.
A calculation with the 
dynamical cluster approximation (DCA) shows that a negative $t_2$ works 
destructively against $d$-wave superconductivity\cite{Maier}, and 
a more realistic DCA calculation that considers the oxygen $p_\sigma$ 
orbitals for the La and Hg cuprates also indicates 
a similar tendency\cite{Kent}.

\subsection{The two-orbital model}

To resolve the above problem for the $d_{x^2-y^2}$ single-orbital 
model, we now 
focus on other orbital degrees of freedom.
In fact, Fig.\ref{fig1} shows that in the La system 
the main band has a strong $d_{z^2}$ character 
around the N point on the Fermi surface that corresponds to the 
wave vectors $(\pi,0),(0,\pi)$ in a square lattice. 
This has been recognized from an 
early stage of  the study on the cuprates
\cite{Shiraishi,Eto,Freeman,DiCastro}, 
and more recently, it has been discussed in refs.\cite{Andersen,Pavarini} 
that the mixture of $d_{z^2}$ character to the main component determines 
the shape of the Fermi surface. 
Namely, the large $d_{z^2}$ contribution in the La system makes the 
Fermi surface closer to a square (i.e., a diamond), 
while in the Hg cuprate the $d_{z^2}$ contribution is small and the 
Fermi surface is more rounded (as confirmed in the following). 

\begin{figure}[!b]
\includegraphics[width=7cm]{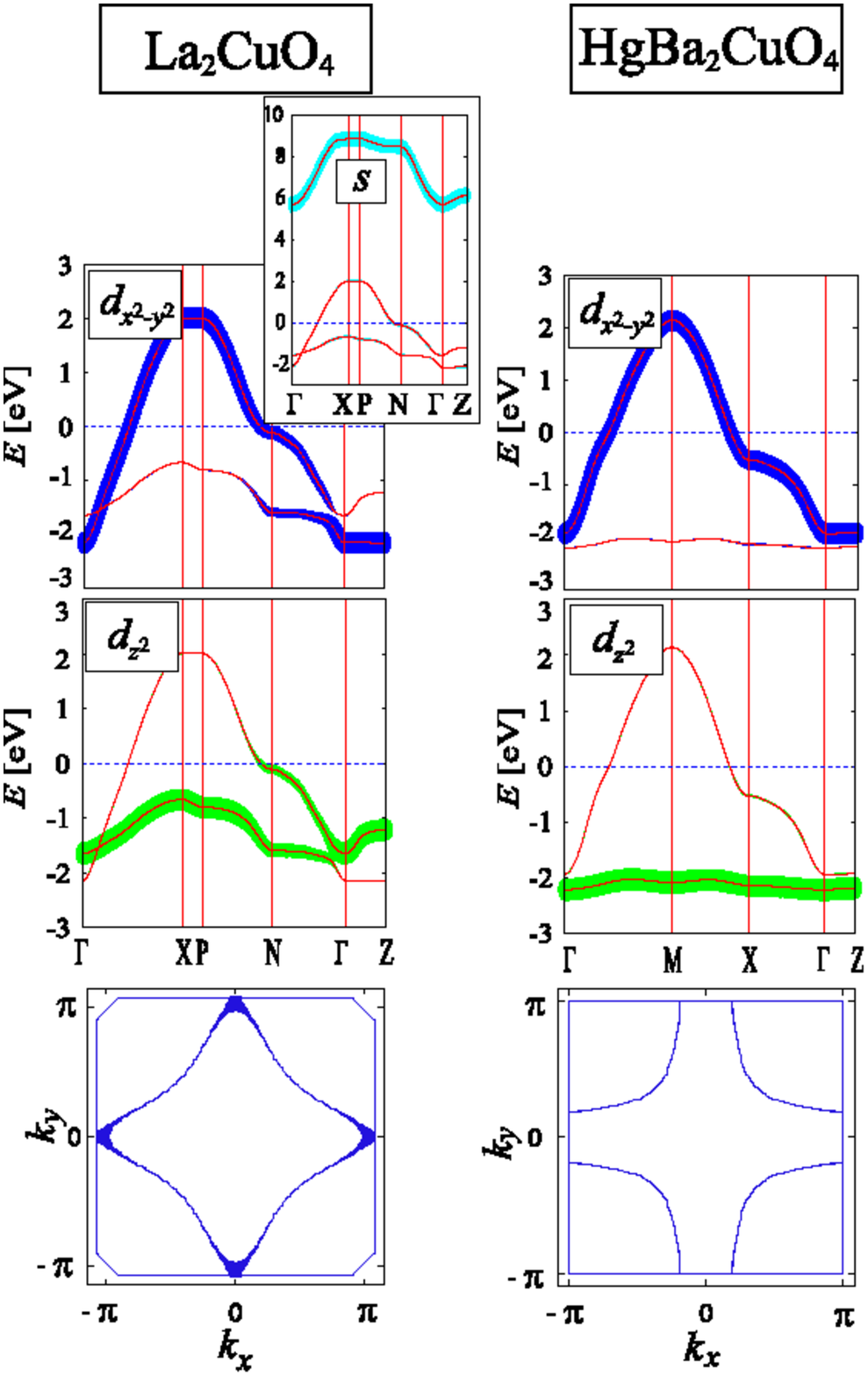}
\caption{(Color online) The band structure
(with $E_F = 0$) 
in the two-orbital ($d_{x^2-y^2}$-$d_{z^2}$) model 
for La$_2$CuO$_4$ (left column) and 
HgBa$_2$CuO$_4$ (right). The top (middle) panels 
depict the weights of the $d_{x^2-y^2}$ 
($d_{z^2}$) characters with thickened lines, while the bottom panels are the 
 Fermi surface for the band filling of $n=2.85$.  
The inset shows the band 
structure of the three-orbital model (see text) for La system, where 
the 4s character is indicated by thick lines.}
\label{fig3}
\end{figure}

In order to understand the experimentally observed 
correlation between the Fermi surface shape and $T_c$, 
we consider a two-orbital model that takes into account not only the 
$d_{x^2-y^2}$ Wannier orbital but also the $d_{z^2}$ Wannier orbital 
explicitly\cite{SakakibaraPRL}.  
A first-principles calculation\cite{wien2k,pwscf} is used to construct 
maximally localized Wannier
orbitals\cite{w2w,MaxLoc}, from which the hopping integrals and the on-site 
energies of the 
two-orbital tight-binding model for the La and Hg cuprates are deduced. 
Thus obtained band structures of the two-orbital model for the La and Hg cuprates 
are shown in Fig.\ref{fig3},  along with 
the Fermi surface for the band filling of $n=2.85$ 
($n=$number of electrons per site), which corresponds to 0.15 holes per 
Cu atom.

\begin{figure}[!h]
\includegraphics[width=7cm]{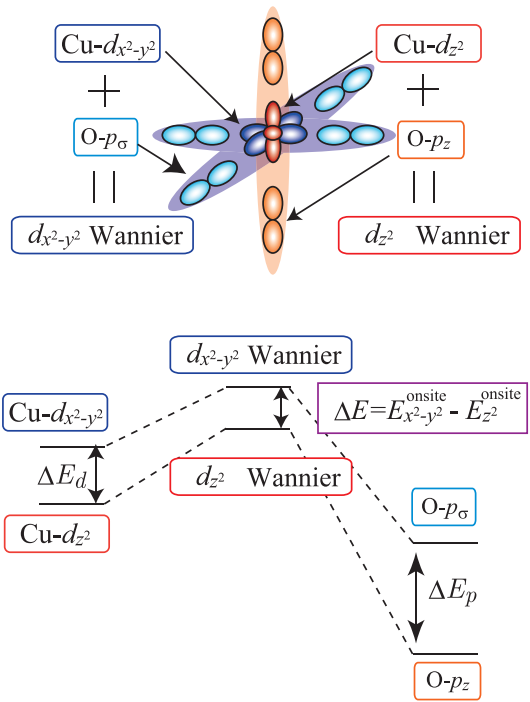}
\caption{(Color online) 
The top panel shows the main components of the two 
Wannier orbitals (having different types of $\sigma$ bonding) 
considered in the present two-orbital model. 
The bottom panel shows the definition of the level offsets $\Delta E$, 
$\Delta E_d$, and $\Delta E_p$.
}
\label{fig4}
\end{figure}

In the present two-orbital model, 
the $d_{x^2-y^2}$ Wannier orbital originates 
primarily from the Cu 3$d_{x^2-y^2}$ 
and the in-plane O 2$p_{\sigma}$ orbitals. 
On the other hand, the $d_{z^2}$ Wannier
orbital originates mainly from the Cu 3$d_{z^2}$  and apical O 2$p_{z}$
orbitals.
Namely, this model incorporates two types of $d$-$p_\sigma$ antibonding
states, where the former spreads over the CuO$_2$-plane while the 
latter along the $c$-axis(Fig.\ref{fig4}). 
Table\ref{table1} shows the parameter values of the present model, 
from which we can identify that  $d_{x^2-y^2}-d_{z^2}$ interorbital
hopping occurs mainly between nearest-neighbor Cu sites, which gives 
rise to the orbital mixture. 
Because the $d_{x^2-y^2}-d_{z^2}$ hopping 
integrals are similar for the La and Hg compounds, 
the onsite energy difference $\Delta E = E _{x^2-y^2}-E_{z^2}$ 
between the two orbitals can be used as a measure of the 
$d_{z^2}$ mixture.  
Note that the interorbital hoppings have different signs between $x$ and
$y$ directions, i.e., the matrix element has the form
$-2t_1[\cos(k_x)-\cos(k_y)]$, 
so that the $d_{x^2-y^2}-d_{z^2}$ mixture is strong 
around the wave vectors $(\pi,0), (0,\pi)$ (N point in the La cuprate), 
while small around $|k_x|=|k_y|$.

In Table\ref{table1} we also show the parameters for the single-orbital model obtained by the similar method.
In the single-orbital model, the ``$d_{x^2-y^2}$'' Wannier 
orbital effectively contains the $d_{z^2}$ orbital in the tail 
parts of the Wannier orbital.

\begin{table}[!h]
\caption{Hopping integrals within the $d_{x^2-y^2}$ orbital  
for the single- and two-orbital models (upper half), 
interorbital hopping (middle), 
and $\Delta E\equiv E_{x^2-y^2}-E_{z^2}$ (bottom).}
\label{table1}
\begin{tabular}{ c| l l l l}
\hline
 & 1-orbital & & \hspace{1.0cm} 2-orbital & \\
 &  La & Hg  &\hspace{1.0cm}  La & Hg\\ \hline
$t(d_{x^2-y^2}\rightarrow d_{x^2-y^2}) $\\ 
$t_1 \rm{[eV]}$ & -0.444 & -0.453 & \hspace{1.0cm} -0.471 & -0.456 \\ 
$t_2 \rm{[eV]}$ & 0.0284 & 0.0874 & \hspace{1.0cm} 0.0932 & 0.0993 \\
$t_3 \rm{[eV]}$ & -0.0357 & -0.0825  & \hspace{1.0cm}  -0.0734 & -0.0897 \\
$(|t_2|+|t_3|)/|t_1|$ & 0.14 &  0.37 & \hspace{1.0cm} 0.35 & 0.41 \\ \hline 
$t(d_{x^2-y^2}\rightarrow d_{z^2})$\\ 
$t_1 \rm{[eV]}$ & - & -&\hspace{1.0cm}  0.178 & 0.105 \\
$t_2 \rm{[eV]}$ & - & -&\hspace{1.0cm}  small & small \\
$t_3 \rm{[eV]}$ & - & -&\hspace{1.0cm}  0.0258 & 0.0149 \\\hline
$\Delta E\rm{[eV]}$ & -  &  - &\hspace{1.0cm}  0.91 & 2.19 \\
\hline
\end{tabular}
\end{table}

\subsection{Correlation between the curvature of the 
Fermi surface and $\Delta E$}   

\begin{figure}[!h]
\includegraphics[width=7cm]{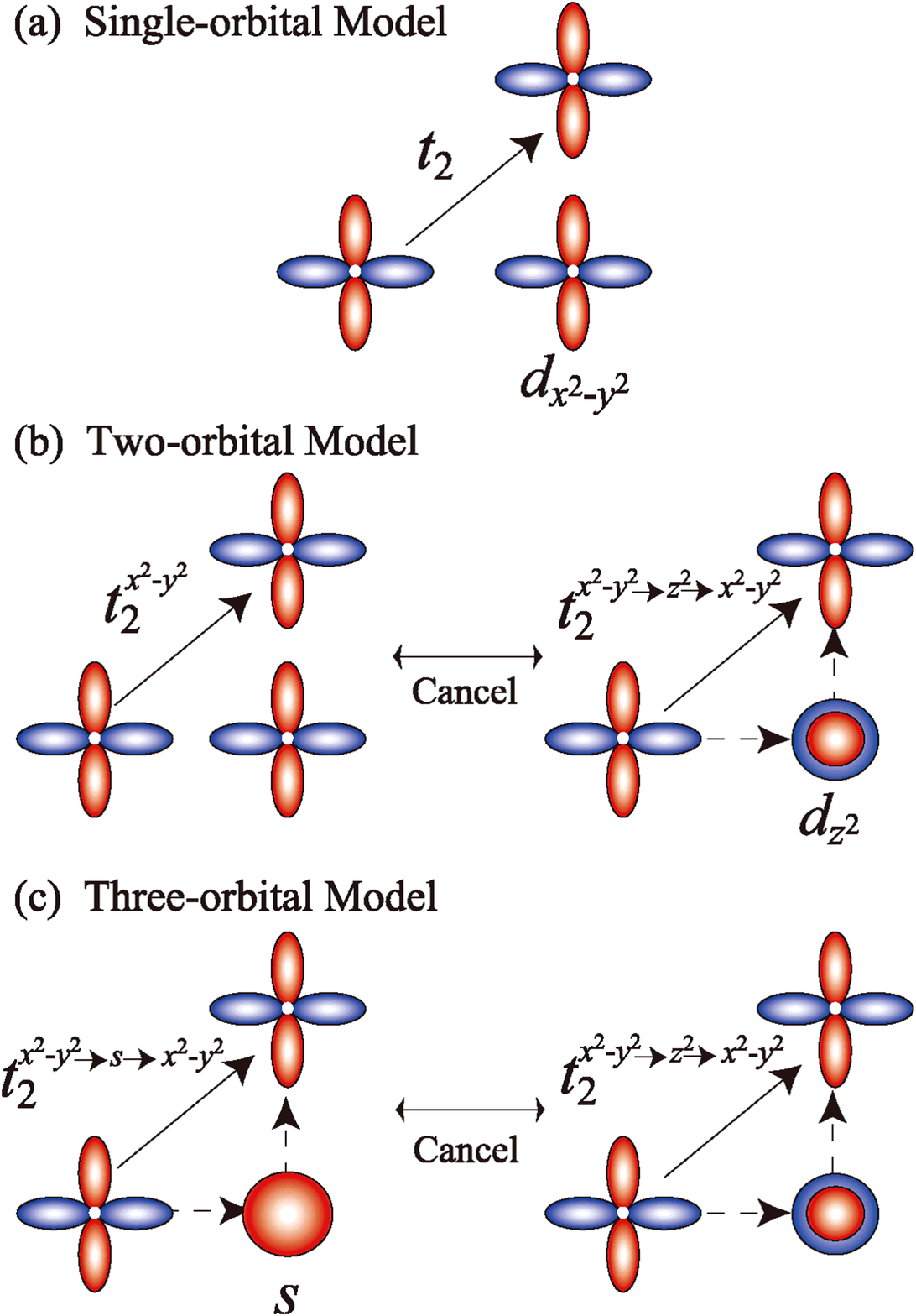}
\caption{(Color online) 
Origin of the effective second-neighbor 
hopping ($t_2$ in the single-band model, (a)) in the two-orbital(b) and 
three-orbital(c) models.
}
\label{fig5}
\end{figure}

The $d_{z^2}$ orbital contribution  
has also a large effect on the curvature of the Fermi surface\cite{Andersen,Pavarini}, which 
can indeed be seen from Table\ref{table1} as follows.
In the single-orbital model, the La cuprate has smaller $t_2$ and $t_3$ 
as compared to the Hg cuprate 
(with the ratio $(|t_2|+|t_3|)/|t_1|$ being 0.14(0.37) for La(Hg)), 
resulting in the smaller curvature 
of the Fermi surface in the former as mentioned. 
On the other hand, in the two-orbital model that considers the 
$d_{z^2}$ orbital explicitly, 
the ratio $(|t_2|+|t_3|)/|t_1|$ within the $d_{x^2-y^2}$ orbital changes to 
0.35(0.41) for the La(Hg).
The value is nearly the same between the single- and two-orbital modelling of Hg, while the value is significantly increased 
in the two-orbital model for La.  
The reason why $t_2$ and $t_3$ in the two-orbital model for La is 
large as compared to that in the single-orbital model can be understood 
from Fig.\ref{fig5} as follows. 
Let us consider the diagonal hopping ($t_2$). 
There is a direct ($d_{x^2-y^2}-d_{x^2-y^2}$) diagonal hopping, but 
there is also an indirect diagonal hopping 
that becomes effective when $\Delta E$ is small, that is, 
$d_{x^2-y^2}\rightarrow d_{z^2}\rightarrow d_{x^2-y^2}$.
In the single-orbital model, 
where the $d_{z^2}$ component is effectively included in 
the $d_{x^2-y^2}$ Wannier orbital, the contribution of the 
$d_{x^2-y^2}\rightarrow d_{z^2}\rightarrow d_{x^2-y^2}$ path is 
effectively included in $t_2$.
The latter contribution has a sign opposite 
to that of the direct diagonal hopping 
(the reason of which will be clarified later), 
so that we end up with a small effective $t_2$ in the single-orbital model 
when $\Delta E$ is small as in the La cuprate.  
A similar argument applies to $t_3$. 
Conversely, the Hg cuprate has a large $\Delta E$ so that 
the $d_{z^2}$ contribution barely exists in the single-orbital model, 
and the ratio $(|t_2|+|t_3|)/|t_1|$ is similar to that in the 
two-orbital model.

In the La cuprate, the $d_{x^2-y^2}$ and $d_{z^2}$ orbitals 
strongly mix around the N point, so that the upper and lower bands 
repel with each other there, and the saddle point of the 
upper band that corresponds to the van Hove singularity is pushed up to 
nearly touch the Fermi level for the band filling of $n=2.85$. 
Thus the Fermi surface almost touches 
the wave vectors $(\pi,0)$, $(0,\pi)$. In the Hg cuprate, there is no 
such splitting of the two bands, and the saddle point stays 
well below the Fermi level, resulting in a rounded Fermi surface that 
is closed around the wave vector $(\pi, \pi)$.

\section{Many-body calculation of the superconductivity}

\subsection{Calculation method}

We now consider a many-body Hamiltonian based on the two-orbital 
tight-binding model discussed above, which is given, in the 
standard notation, as 
\begin{eqnarray}
H &=& \sum_i\sum_\mu\sum_{\sigma}\varepsilon_\mu n_{i\mu\sigma}
 + 
\sum_{ij}\sum_{\mu\nu}\sum_{\sigma}t_{ij}^{\mu\nu}
c_{i\mu\sigma}^{\dagger} c_{j\nu\sigma}\nonumber\\
&+&\sum_i\left( U\sum_\mu n_{i\mu\uparrow} n_{i\mu\downarrow}
+U'\sum_{\mu > \nu}\sum_{\sigma,\sigma'} n_{i\mu\sigma} n_{i\mu\sigma'}
\right.\nonumber\\
&-&\frac{J}{2}\sum_{\mu\neq\nu}\sum_{\sigma,\sigma'} c_{i\mu \sigma}^\dagger c_{i\mu \sigma'} 
c_{i\nu \sigma'}^\dagger c_{i\nu \sigma}\nonumber\\
&+&\left. J'\sum_{\mu\neq\nu} c_{i\mu\uparrow}^\dagger c_{i\mu\downarrow}^\dagger
c_{i\nu\downarrow}c_{i\nu\uparrow}
\right), 
\end{eqnarray}

where $i,j$ denote the sites and $\mu,\nu$ the two-orbitals, while 
the electron-electron interactions comprise 
the intraorbital repulsion $U$, interorbital repulsion 
$U'$, and the Hund's coupling $J$(= pair-hopping interaction $J'$ ).  
Here we take $U=3.0$ eV, $U'=2.4$ eV, and 
$J$=0.3 eV\cite{comment3}.  
These values conform to a widely accepted, first-principles estimations  for the cuprates that the $U$ is 7-10t 
(with $t \simeq$ 0.45 eV), while $J, J' \simeq 0.1U$.   
Here we also observe the orbital SU(2) requirement $U'=U-2J$.  

To study the superconductivity in this multi-orbital 
Hubbard model, we apply FLEX approximation\cite{Bickers,Dahm,Kontani}.
In FLEX, we start with the Dyson's equation to obtain 
the renormalized Green's function, which, in the multi-orbital case, 
is a matrix in the orbital 
representation as $G_{l_1l_2}$, where $l_1$  and $l_2$ are orbital indices.
The bubble and ladder diagrams consisting of 
the renormalized Green's function are then summed to obtain the 
spin and charge susceptibilities, 
\begin{equation}
\hat{\chi}_s(q)=\frac{\hat{\chi}^0(q)}{1-\hat{S}\hat{\chi}^0(q)} ,
\end{equation}
\begin{equation}
\hat{\chi}_c(q)=\frac{\hat{\chi}^0(q)}{1+\hat{C}\hat{\chi}^0(q)} ,
\end{equation}
where $q \equiv (\Vec{q},i\omega_n)$, 
the irreducible susceptibility is 
\begin{equation} 
\chi^0_{l_1,l_2,l_3,l_4}(q) =\sum_q G_{l_1l_3}(k+q)G_{l_4l_2}(k),
\end{equation}
with the interaction matrices 
\begin{equation}
S_{l_1l_2,l_3l_4}
=\left\{\begin{array}{cc}
U, &\;\; l_1=l_2=l_3=l_4\\ 
U',&\;\; l_1=l_3\neq l_2=l_4\\
J,&\;\; l_1=l_2\neq l_3=l_4\\
J',&\;\; l_1=l_4\neq l_2=l_3 ,
\end{array}  \right.
\end{equation}

\begin{equation}
C_{l_1l_2,l_3l_4}
=\left\{\begin{array}{cc}
 U &\;\; l_1=l_2=l_3=l_4\\ 
-U'+J & \;\; l_1=l_3\neq l_2=l_4\\
2U'-J,&\;\; l_1=l_2\neq l_3=l_4\\
J'& \;\; l_1=l_4\neq l_2=l_3 .
\end{array}  \right.
\end{equation}
With these susceptibilities, the fluctuation-mediated 
effective interactions are obtained, which are used to calculate the 
self-energy. Then the renormalized 
Green's functions are determined self-consistently from the Dyson's
equation. 
The obtained Green's functions and the susceptibilities are 
used to obtain the spin-singlet pairing interaction in the form
\begin{equation}
\hat{V}^s(q)=\frac{3}{2}\hat{S}\hat{\chi}_s(q)\hat{S}
-\frac{1}{2}\hat{C}\hat{\chi}_c(q)\hat{C}+\frac{1}{2}(\hat{S}+\hat{C}),
\end{equation}
and this is plugged into the linearized Eliashberg equation,  
\begin{eqnarray}
\lambda \Delta_{ll'}(k)&=&-\frac{T}{N}\sum_q
\sum_{l_1l_2l_3l_4}V_{l l_1 l_2 l'}(q)\nonumber\\
&\times& G_{l_1l_3}(k-q)\Delta_{l_3l_4}(k-q)
G_{l_2l_4}(q-k).
\end{eqnarray}
The superconducting transition temperature, $T_c$, 
corresponds to the temperature at which the eigenvalue $\lambda$ 
of the Eliashberg equation reaches unity, 
so that $\lambda$ at a fixed temperature can be used as a measure
for $T_c$.  
In the present calculation, the temperature is 
fixed at $k_{\rm B}T= 0.01$ eV, which  amounts to about 100K, 
and the band 
filling (number of electrons/site) is set to be $n=2.85$, 
which corresponds to 0.85 electrons per site in the main band, 
namely, around the optimum doping concentration.  
We take  $32\times 32 \times 4$ $k$-point 
meshes and 1024 Matsubara frequencies.

\subsection{Correlation between $T_c$ and $\Delta E$}
\label{IIIB}

\begin{figure}[!b]
\includegraphics[width=7cm]{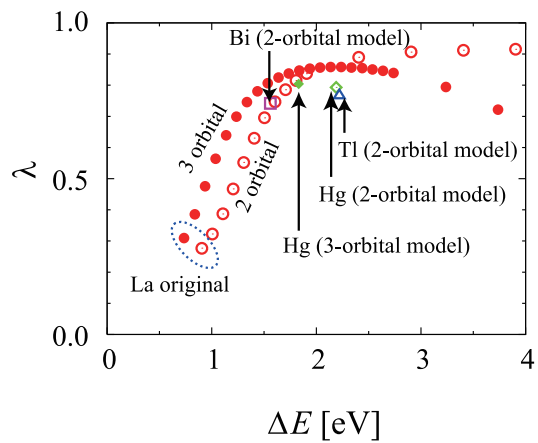}
\caption{ (Color online) The eigenvalue, $\lambda$, of the Eliashberg equation for $d$-wave superconductivity plotted against 
$\Delta E=E_{x^2-y^2}-E_{z^2}$ 
for the two-orbital (red open circles) and three-orbital (red filled circles)
models for La$_2$CuO$_4$. 
Corresponding eigenvalues 
for HgBa$_2$CuO$_4$,Bi$_2$Sr$_2$CuO$_6$ and Tl$_2$Ba$_2$CuO$_6$ are also indicated.
}
\label{fig6}
\end{figure}

Let us now  investigate how the $d_{z^2}$ orbital affects superconductivity 
by hypothetically varying $\Delta E$  from its original value 0.91 eV 
(shown in Table\ref{table1}) to 4.0 eV for the La cuprate to 
single out the effect of $\Delta E$.  
The eigenvalue of the Eliashberg equation $\lambda$ 
calculated as a function of $\Delta E$ 
in Fig.\ref{fig6} shows that $\lambda$ initially increases 
rapidly upon increasing $\Delta E$, 
then saturates for $\Delta E> 3$eV.
This means that the mixture of the 
$d_{z^2}$ orbital on the Fermi surface around the wave vectors 
$(\pi,0)$, $(0,\pi)$ does indeed strongly suppress 
superconductivity in the original La system, while for large enough $\Delta E$ the system essentially reduces to a single-orbital model, 
where $d_{z^2}$ orbital no longer affects superconductivity.  
As mentioned above, the $d_{z^2}$ orbital mixture makes the Fermi surface 
more square shaped, which in itself favors 
superconductivity as mentioned in section\ref{IIA} 
(e.g., Fig.1 of ref.\onlinecite{SakakibaraPRL}).  
Thus we can see that the effect of the 
$d_{z^2}$ orbital mixture {\it supersedes} the effect of 
Fermi surface shape, and $T_c$ is primarily determined 
by the former.  This explains why we have  
$T_c$  positively 
correlated with $\Delta E$ simultaneously with 
the roundness of the Fermi surface that is 
also  positively correlated with $\Delta E$.  
This should lead to the 
experimentally observed correlation between the shape of the Fermi
surface and $T_c$\cite{Pavarini,Tanaka}.

\subsection{Effects of the inter-orbital electron-electron interaction}

Thus the next important question is: why does the mixture of the 
$d_{z^2}$ orbital on the Fermi surface suppress 
superconductivity?  
To investigate the origin, we have varied the interaction values to 
examine the strength of the spin fluctuations and the 
superconducting instability. 
The strength of the spin fluctuation is measured by the 
antiferromagnetic Stoner factor, which, for a multiband system, 
corresponds to the largest eigenvalue of the matrix $\hat{S}\chi_0$.

In the result in Table\ref{table2} 
we can compare the cases for $U'=0$ eV and $U'=2.4$ eV, 
which shows that 
the strength of the spin fluctuation becomes smaller when $U'$ is turned off.
This should be  because $U'$ hinders four electrons 
(two $d_{z^2}$ and two $d_{x^2-y^2}$) to come on the same site.  
Despite this, it can be seen that $\lambda$ is not 
much affected by $U'$, probably because the suppression 
of superconductivity due to the increased charge/orbital 
fluctuations (which is unfavorable for singlet $d$-wave pairing) 
and the enhancement due to the increased 
spin fluctuations roughly cancel with each other.  
We have also examined how the Hund's coupling $J$ affects superconductivity.  
A comparison between $J=0$ and $J=0.3$ 
shows that superconductivity is slightly suppressed when we turn on $J$, 
which is consistent with an observation that the Hund's coupling 
tends to suppress spin-singlet pairing.  
Nevertheless, the effect of $J$ is 
overall small. The conclusion here is that the effect of the interorbital 
interactions on superconductivity is small, so that the main origin of 
the suppression of superconductivity is the mixture of the 
$d_{z^2}$ orbital on the Fermi surface, which will be 
elaborated in the next subsection.

\begin{table}[!h]
\caption{FLEX result for the eigenvalue of the 
Eliashberg equation $\lambda$, and the Stoner factor for various 
values of the interorbital interactions $U'$ and $J$, for fixed 
$U=3.0$ eV and $J'=0.30$ eV}
\label{table2}
\begin{tabular}{c c c c c}
\hline
\hspace{0.5cm} $U'$[eV] \hspace{0.5cm} & \hspace{0.5cm} $J$[eV] \hspace{0.5cm}& \hspace{0.5cm} Stoner \hspace{0.5cm} & \hspace{0.5cm} $\lambda$ \hspace{0.5cm}  \\
2.4 & 0.3  & 0.979 & 0.279 \\
2.4 & 0.0  & 0.978 & 0.335 \\
0.0 & 0.3  & 0.925 & 0.291 \\
0.0 & 0.0  & 0.958 & 0.309 \\
\hline
\end{tabular}
\end{table}

\subsection{Origin of the suppression of 
superconductivity by the $d_{z^2}$ mixing}
\label{IIID}

\begin{figure}[!h]
\includegraphics[width=7cm]{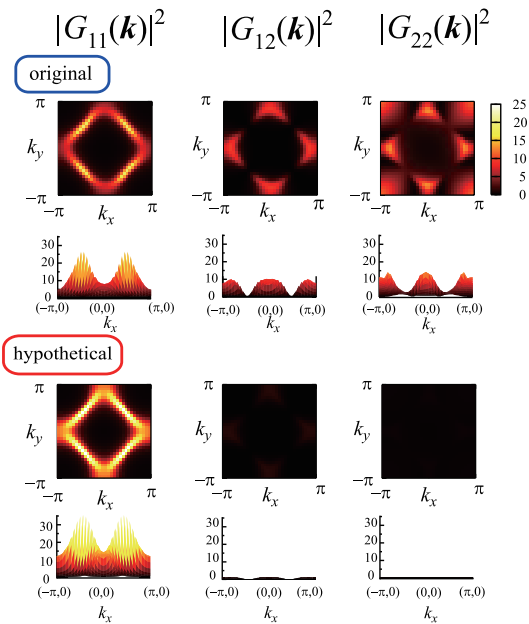}
\caption{(Color online) 
Contour plots and side views of the diagonal and off-diagonal elements of the 
squared Green's function for the original La and the hypothetical cases. 
The subscripts 1 and 2 stand for the $d_{x^2-y^2}$ and $d_{z^2}$ orbitals, respectively.
}
\label{fig7}
\end{figure}

\begin{figure}[!h]
\includegraphics[width=7cm]{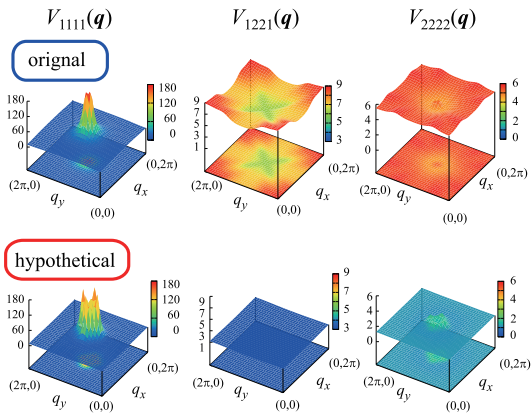}
\caption{(Color online) 
Diagonal and off-diagonal elements of the 
pairing interaction depicted against $(q_x, q_y)$. 
}
\label{fig8}
\end{figure}

Here we pinpoint why the $d_{z^2}$ orbital component 
mixture degrades $d$-wave superconductivity. 
In Fig.\ref{fig7}, we show 
the  squared orbital diagonal and off-diagonal elements of the 
Green's function matrix spanned by the orbital indices 
at the lowest Matsubara frequency.
We compare them for two cases; the original La cuprate and 
a hypothetical case where we increase $\Delta E$ 
to the value for Hg, where the hopping integrals are tuned 
to retain the shape of the Fermi surface to that of the
La cuprate.  In the hypothetical case, the interaction 
values are reduced ($U=2.1$ eV, $J=J'=0.1U$ and $U'=U-2J$) so as to make the maximum value of the pairing interaction in the 
$d_{x^2-y^2}$ channel  ($V_{1111}$) 
to be roughly the same as that in the 
original La case. Then the eigenvalues of the Eliashberg equation 
at $T=0.01$ differ as much as 
$\lambda=0.28$  and 0.88 for the original La and the hypothetical cases, 
respectively.   
Let us analyze the origin of this difference. 
Here we denote the $d_{x^2-y^2}$ and $d_{z^2}$ orbitals 
as orbitals 1 and 2, respectively.
In the original La, compared to the hypothetical case, 
(i) the $d_{x^2-y^2}$ diagonal element $|G_{11}|^2$ is smaller 
especially around the wave vectors $(\pi,0)/(0,\pi)$,  
(ii) the $d_{z^2}$ diagonal element $|G_{22}|^2$ 
is much larger, and  (iii) there 
is a substantial off-diagonal element $|G_{12}|^2$ due to the 
strong $d_{z^2}$ orbital mixture.
If we turn to the pairing interaction matrix, again at the lowest
Matsubara frequency, in Fig.\ref{fig8}, 
the diagonal elements have similar maximum values between the two cases 
because the interaction 
is reduced in the hypothetical one as mentioned above.
In the original La,  the off-diagonal element of the pairing 
interaction $V_{1221}$ is 
large compared to the hypothetical case, and the 
interaction is broadly peaked around $(0,0)$.
On the other hand, the $d_{z^2}$ diagonal interaction $V_{2222}$ 
is finite but has a small momentum dependence.
Considering the above, the dominant contributions to the 
Eliashberg equation regarding the $d_{x^2-y^2}$ orbital component of the 
gap function $\Delta_{11}$ is roughly given as 
\begin{eqnarray}
&&\lambda \Delta_{11}(k)\sim \nonumber\\ 
&-&V_{1111}(Q)G_{11}(k-Q)\Delta_{11}(k-Q)
G_{11}(Q-k)\nonumber\\
&-&V_{1221}(0,0)G_{21}(k)\Delta_{11}(k)
G_{21}(-k)\nonumber\\
&-&\sum_{q} V_{2222}(q)G_{22}(k-q)\Delta_{22}(k-q)
G_{22}(q-k),
\end{eqnarray}
where $\Vec{Q}=(\pi,\pi)$. 
If we consider a wave vector $\Vec{k}$ near $(\pi,0)$ 
on the Fermi surface that 
has a positive $\Delta_{11}(k)$, 
$\Delta_{11}(k-Q)$ will be negative for the $d$-wave gap.
Then the first term on the right-hand side will be positive 
but small in the 
original La compared to the hypothetical case 
because of the small $G_{11}$ especially around $(\pi,0)/(0,\pi)$. 
This is the main reason why $\lambda$ is reduced in the 
original La compared to the hypothetical case.
In addition, the second term, which cannot be neglected when the 
$d_{z^2}$ mixture is significant, actually has a negative sign, 
and also acts to suppress $\lambda$ and hence $T_c$.
The interaction $V_{2222}$ has small momentum dependence,  
so that this term has small contribution for a $d$-wave gap when 
summed over $q$.

In the above comparison, we have reduced the interactions in the 
hypothetical case so as to make the maximum pairing
interaction $V_{1111}$ nearly the same 
as in the original La.
The reason we fix the strength of the pairing interaction is 
because the maximum value of the 
pairing interaction actually does not differ very much 
upon increasing $\Delta E$
in the results given in Fig.\ref{fig6}.
The reason for this, 
despite the Fermi surface nesting becoming 
worse as we increase $\Delta E$, is mainly two-fold: 
(i) the $d_{z^2}$ orbital mixture on the Fermi surface becomes weaker,  
and (ii) inclusion of the self-energy in the FLEX 
weakens the role of the Fermi surface nesting played in the 
development of the spin fluctuations. Regarding the second point, 
in the random phase approximation where the self-energy is not 
considered, the Fermi surface nesting effect 
on the strength of the spin fluctuations, 
hence the pairing interaction, 
is so strong that $\lambda$ does not increase with 
$\Delta E$ as in Fig.\ref{fig6} (and thus the $T_c$ difference 
between La and Hg cuprates discussed later cannot be explained), 
although the effect of the increase in $G_{11}$  due to the 
reduction of the $d_{z^2}$ mixture is present.  
This may be regarded as consistent with a recent result obtained with 
the functional renormalization group, where the 
self-energy correction is not considered\cite{Honerkamp}.

\subsection{$d_{x^2-y^2}+d_{z^2}+s$ three-orbital model}

So far we have analyzed the two-orbital model that considers the 
$d_{x^2-y^2}$ and $d_{z^2}$ Wannier orbitals.
Actually in ref.\cite{Andersen,Pavarini}, it has been pointed out 
that the "axial state" that contains not only Cu $d_{z^2}$ and 
O$_{\rm{apical}} p_z$ orbital but also the Cu-$4s$ orbital 
is important in determining the shape of the Fermi surface.
In the present two-orbital model, the Cu $4s$ orbital is 
effectively incorporated in both of the $d_{x^2-y^2}$ and $d_{z^2}$
Wannier orbitals. Namely, the Wannier orbitals have Cu $4s$ 
components in their tails.  
In order to examine the effect of Cu $4s$ orbital 
more explicitly, let us 
consider in this section a $d_{x^2-y^2}+d_{z^2}+s$ 
three-orbital model which takes into account the Cu-$4s$ Wannier 
orbital on an equal footing.

In this model, the $4s$ Wannier orbital is a mixture mainly of 
Cu $4s$ and O $p_{\sigma}$ orbitals.
The O $p_{\sigma}$ orbitals contain not 
only the in-plane O$_{\rm{plane}} p_{\sigma}$ but also 
the apical O$_{\rm{apical}} p_z$.
The main band originating from the 4s orbital for the La system is shown in 
the inset of Fig\ref{fig3}.   While the 
4$s$ band lies well($\simeq 7$ eV) above the Fermi level, 
the $4s$ orbital still gives an important 
contribution to the Fermi surface shape.
Here again we estimate the ratio $(|t_2|+|t_3|)/|t_1|$ 
within the $d_{x^2-y^2}$, where we find a much smaller value of 
$0.10$ against $0.35$ in the two-orbital model. 
This means that the large $t_2$ and $t_3$ 
within the $d_{x^2-y^2}$ Wannier orbital in the two-orbital model 
is mainly due to the 
$d_{x^2-y^2}\rightarrow 4s \rightarrow d_{x^2-y^2}$ hopping path 
(Fig.\ref{fig5}, bottom panel), 
as pointed out in ref.\cite{Pavarini}.  
Then, from the viewpoint of the three-orbital model, 
$t_2$ and $t_3$ in the 
single-orbital model of La cuprate are small because the 
$d_{x^2-y^2}\rightarrow 4s \rightarrow d_{x^2-y^2}$ and 
$d_{x^2-y^2}\rightarrow d_{z^2}\rightarrow d_{x^2-y^2}$ 
contributions nearly cancel with each other.  
The two effective hoppings have opposite signs 
because the $d_{z^2}$ level lies below $d_{x^2-y^2}$ 
while $4s$ lies above.

Now we apply FLEX to this three-orbital model, where we vary 
$\Delta E = E _{x^2-y^2}-E_{z^2}$ 
and calculate the eigenvalue of the Eliashberg equation 
as we did in subsection\ref{IIIB}.
Here, we fix the on-site 
energy difference $E_s-E_{d_{z^2}}$ at its original value when we 
vary $\Delta E$, 
because the three-orbital model for the 
Hg compound has roughly the same $E_s-E_{d_{z^2}}$ 
as that of the  La compound.

The result is displayed in Fig. \ref{fig6} as marked with "3-orbital".  
We recognize that in the small $\Delta E$ regime the eigenvalue
$\lambda$ rapidly increases with $\Delta E$ 
as in the two-orbital model.  
In the large $\Delta E$ regime, however, 
$\lambda$ tends to decrease 
rather than to saturate.  
In this regime, the $4s$ level comes too close to the Fermi level, 
and strongly deforms the Fermi surface.  
Nonetheless, considering that even in the case of the 
Hg compound with a larger $\Delta E$ as will be discussed later, 
$\Delta E$ (3-orbital model) is still $\simeq 2$ eV, 
i.e., 
such a suppression of superconductivity due to the 
$4s$ level coming too close to the Fermi level is 
not expected in real materials.

Thus we can conclude on the $4s$ orbital that, while this 
orbital has an important effect on the shape
of the Fermi surface, the effect can be included in the two-orbital 
model,  so that the FLEX results for the two- and three-orbital 
models are similar 
as far as the $T_c-\Delta E$ relation is concerned 
(unless we consider unrealistically large $\Delta E$).  
This is natural in that the level offset 
$E_{x^2-y^2}-E_{z^2}$ is smaller ($\simeq 1$ eV) than 
the electron-electron interaction($\simeq 3$ eV), 
while the $E_{s}-E_{x^2-y^2}$ is much larger ($\simeq 7$ eV).  
Hence the 4$s$ orbital can effectively be integrated out 
before the many-body analysis, while the $d_{z^2}$ orbital cannot.  
In this sense the two-orbital($d_{x^2-y^2}-d_{z^2}$) model suffices 
for discussing the material dependence of the $T_c$ in the cuprates.

\section{Material dependence of $\Delta E$}

\begin{figure}[!h]
\includegraphics[width=7cm]{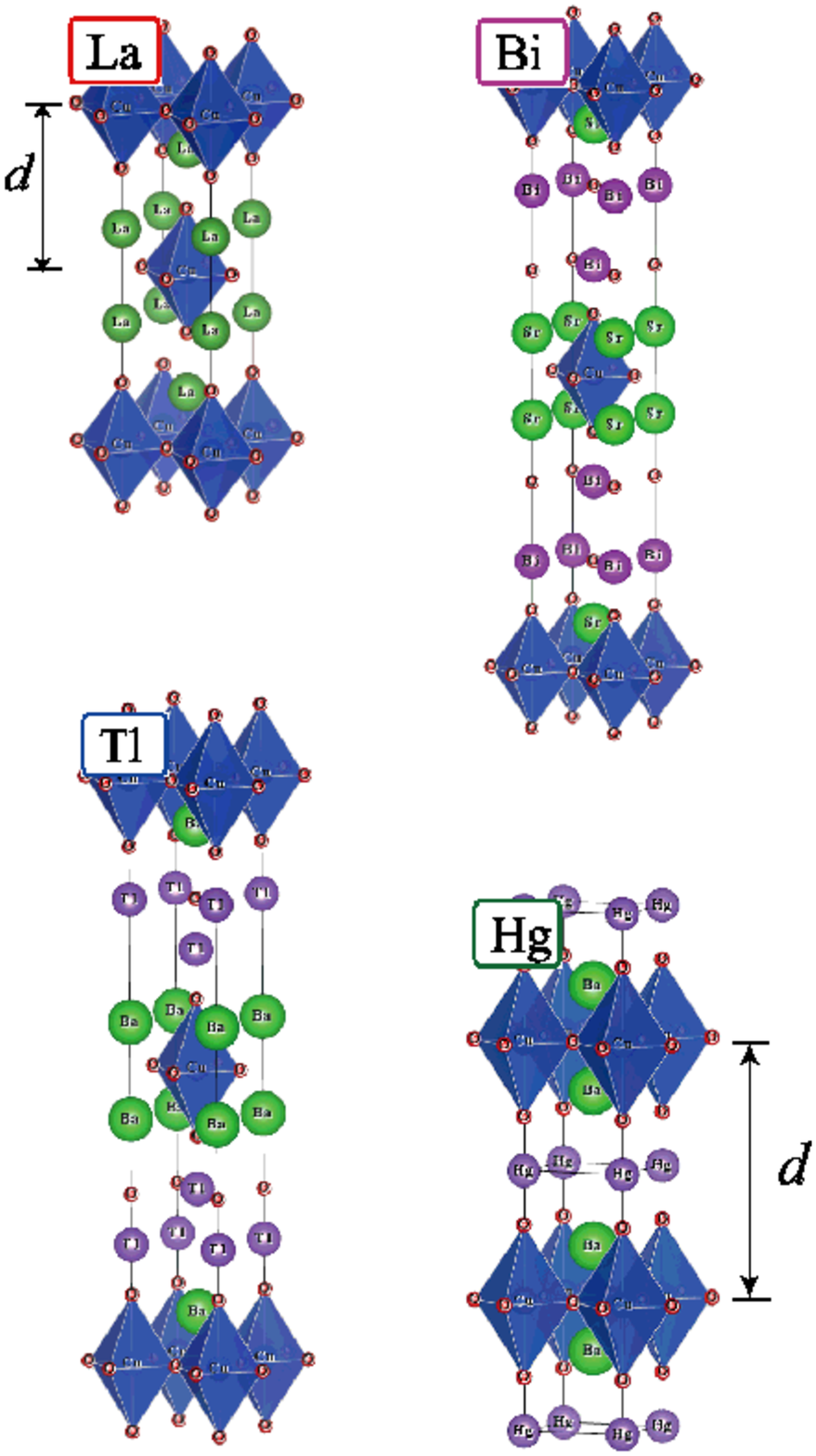}
\caption{(Color online) 
Lattice structures of La$_2$CuO$_4$, Bi$_2$Sr$_2$CuO$_6$, Tl$_2$Ba$_2$CuO$_6$, and HgBa$_2$CuO$_4$.
}
\label{fig9}
\end{figure}

We have seen that the mixture of the $d_{z^2}$ component 
strongly affects superconductivity, making $T_c$ positively correlated with 
$\Delta E$.
To further endorse this, 
we have plotted in Fig.\ref{fig6} the eigenvalue $\lambda$ for 
the two-orbital models for single-layered cuprates 
Bi$_2$Sr$_2$CuO$_6$\cite{Bi-st}, Tl$_2$Ba$_2$CuO$_6$\cite{Tl-st}, and HgBa$_2$CuO$_4$ as well, 
whose lattice structures are shown in Fig.\ref{fig9}.  
We can see that these materials also fall upon 
reasonably well on the correlation between $\lambda$ and $\Delta E$.  
Thus the next fundamental question 
in understanding the material dependence of $T_c$ is: which 
key factors determine $\Delta E$.  
This section precisely addresses that question.

\subsection{Crystal-field effect}

Since the main components of the Wannier orbitals in the 
two-orbital model are the 
Cu $3d_{x^2-y^2}$ and Cu $3d_{z^2}$ orbitals, the crystal-field 
splitting between 
these orbitals, denoted as $\Delta E_d$ here, 
should be the first key factor governing $\Delta E$.
Namely, materials with a larger apical oxygen height above the 
CuO$_2$ plane($h_{\rm{O}}$) should have 
a larger crystal-field splitting,\cite{Eto} 
so that $\Delta E_d$, and thus $\Delta E$, should be larger (Fig.\ref{fig4}).
Indeed, the La compound has smaller $h_{\rm{O}}=2.41$ \AA\   
and $\Delta E$, while the Hg compound has 
larger $h_{\rm{O}}=2.78$ \AA\  and $\Delta E$.

\begin{figure}[!h]
\includegraphics[width=7cm]{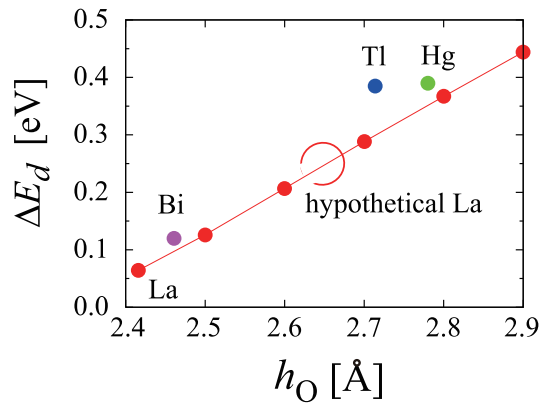}
\caption{(Color online) 
$\Delta E_d$ plotted against $h_{\rm O}$. Solid (red) circles 
connected by a line is the 
result for the hypothetical lattice structure of La cuprate, while 
values for 
Bi, Tl and Hg cuprates are also shown.
}
\label{fig10}
\end{figure}

So let us first focus on how the apical oxygen height $h_{\rm O}$ 
affects $\Delta E_d$. Namely, we construct a model that considers all of
the Cu $3d$ and O $2p$ orbitals(five $3d$ + $3\times4$ $2p =17$ orbitals) explicitly, exploiting 
maximally 
localized Wannier orbitals, and then estimate the on-site energy
difference  between Cu $d_{x^2-y^2}$ and Cu $d_{z^2}$ orbitals as 
$\Delta E_d$. We note that this $\Delta E_d$ is something different from 
$\Delta E$ defined for the effective two-orbital model we have 
considered, since we now 
explicitly consider the oxygen $2p$ orbitals. 
In Fig.\ref{fig10}, we plot $\Delta E_d$ as a function of
$h_{\rm O}$, where we hypothetically vary the height for the La system 
from its original value 2.41 \AA\  to 2.90 \AA\ .  
The result shows that $\Delta E_d$ and $h_{\rm O}$ are linearly correlated. 
We have also constructed similar $d$-$p$ models for the Bi, Tl and Hg systems, 
and we can see that the $\Delta E_d$ values for these materials, 
also included in the figure, roughly fall upon the 
linear correlation for the hypothetical La system, which 
indicates that 
$\Delta E_d$ is primarily determined by $h_{\rm O}$.
Such a correlation has also been found in a recent 
quantum chemical calculation\cite{Fulde}, where the $d_{x^2-y^2}$-$d_{z^2}$ 
level splitting evaluated there corresponds more closely to 
the present $\Delta E_d$ rather than $\Delta E$.

\begin{figure}[!h]
\includegraphics[width=7cm]{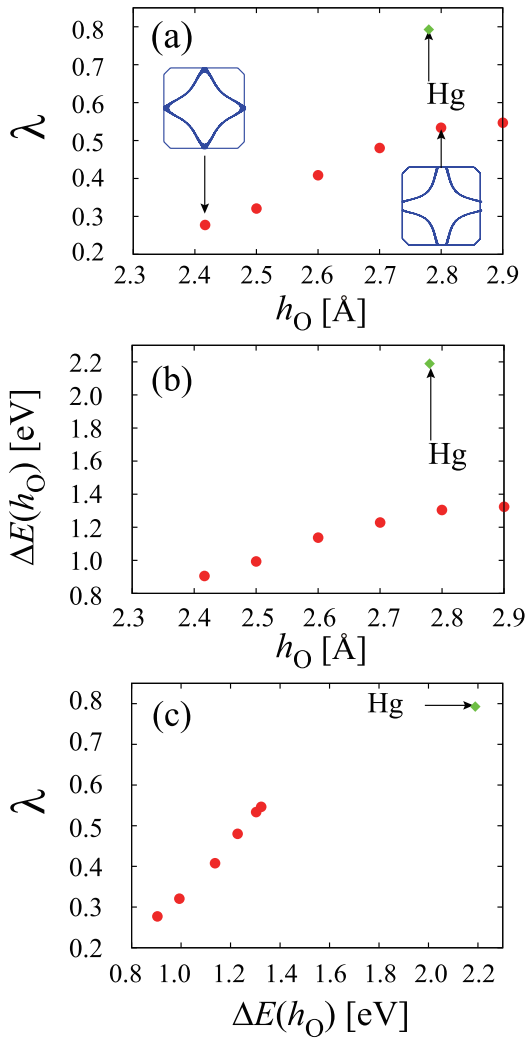}
\caption{(Color online) 
The eigenvalue of the Eliashberg equation 
$\lambda$ (circles) when $h_{\rm O}$ is varied (a) 
or $\Delta E(h_{\rm O})$ is varied (c) 
hypothetically in the lattice structure of La cuprate.
Also plotted is $\Delta E(h_{\rm O})$ against $h_{\rm O}$(b). 
Diamonds in green indicate the values for HgBa$_2$CuO$_4$.
}
\label{fig11}
\end{figure}

Having seen that $h_{\rm O}$ governs $\Delta E_d$, 
we next look at $\Delta E$ and the eigenvalue $\lambda$ 
in the two-orbital models for the La cuprate with 
hypothetically varied $h_{\rm{O}}$. 
As expected, $\Delta E$ in Fig.\ref{fig11} (b) monotonically increases with 
$h_{\rm O}$.  Then $\lambda$ 
(Fig.\ref{fig11} (a)) increases with $h_{\rm O}$, which is in accord with 
the positive correlation between 
$\Delta E$  and $\lambda$ discussed above (Fig.\ref{fig11} (c)). 
Thus $h_{\rm O}$ is shown to be one of the key 
parameters that determine $\Delta E$ and thus $T_c$.

However, if we plot the corresponding values for the 
Hg cuprate, also displayed in the figure, 
we find that $\Delta E$, and thus $\lambda$,  
are larger than those for the 
hypothetical La cuprate for the same apical oxygen height between 
the two cuprates.  
This implies that $h_{\rm{O}}$ and $\Delta E_d$ are 
not the sole parameters that 
determine $\Delta E$ and hence $T_c$, and another factor 
should be lurking.

\subsection{Oxygen-orbital effects}

The above observation has motivated us to 
look more closely into the effects of oxygen orbitals.  
As shown in Fig.\ref{fig4}, the Wannier orbitals in our  
two-orbital model 
the Cu-$3d_{x^2-y^2}$ and $3d_{z^2}$ orbitals strongly hybridize 
with the in-plane O 2$p_\sigma$ and apical oxygen O 2$p_z$ orbitals, 
respectively.  
Thus we can surmise that 
$\Delta E$ should also be affected by the energy difference 
(denoted as $\Delta E_p$) 
between the in-plain $p_{\sigma}$ and the apical oxygen $p_z$.  
By definition, one can expect that $\Delta E_p$ is positively 
correlated with $\Delta V_A$, the 
Madelung potential difference between O$_{\rm plane}$ and 
O$_{\rm apical}$ introduced by 
Ohta {\it et al.} as an important parameter that controls the 
material dependence of the $T_c$\cite{Maekawa}.  
In fact, $\Delta V_A$ for Hg is about 7 eV larger than that of La,
namely, the O-$2p_z$ energy level with respect to the in-plane 
O-2$p_\sigma$ level is much lower in Hg.  

The difference mainly comes
from the crystal structure where 
the apical oxygen in the La cuprate is surrounded by 
other apical oxygens belonging to the neighboring layers, 
while in Hg those oxygen atoms are much further apart 
as seen in Fig.\ref{fig9}.  
This gives a clue to understand the reason why  the hypothetical 
La cuprate with the same $h_{\rm O}$ as  Hg has smaller 
$\Delta E$ and $\lambda$; although $\Delta E_d$ is 
similar between the two systems, $\Delta E_p$ very much differs.
Thus the difference between La and Hg can be attributed to
the distance between neighboring CuO$_2$ layers that is affected by 
the lattice structure, i.e., body-centered tetragonal (bct) vs. simple 
tetragonal.  
However, a similar variance in the layer distance can occur even within 
similar lattice structures.  
La, Bi, and Tl compounds all have the bct structures, 
so naively one might expect similar values of $\Delta V_A$.  However, 
$\Delta V_A$'s for Bi and Tl are much larger than that for La.  
This is because in Bi (Tl) there is a Bi-O (Tl-O) layer 
inserted between the adjacent CuO$_2$ layers (see Fig.\ref{fig9}), 
resulting in a large CuO$_2$ layer separation.

\begin{figure}[!h]
\includegraphics[width=7cm]{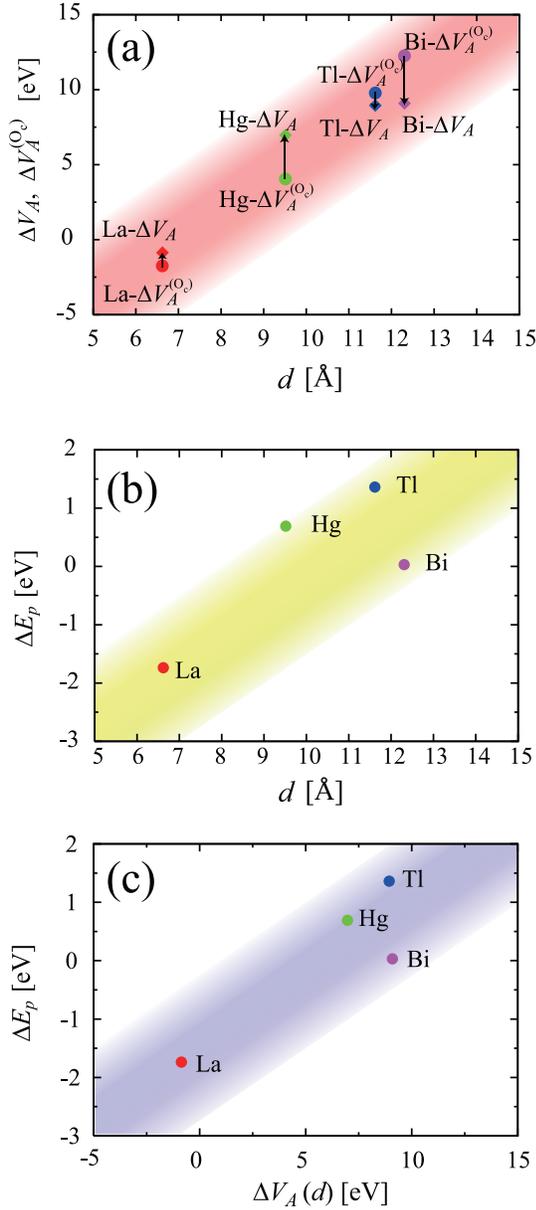}
\caption{(Color online) 
(a) $\Delta V_A^{({\rm O}_c)}$ (circles) and 
$\Delta V_A$ (diamonds) plotted against the layer separation 
$d$ for La, Bi,  Hg, and Tl cuprates.  
(b) The level offset, $\Delta E_p$, between the in-plain $p_{\sigma}$ and the apical oxygen $p_z$ against the layer separation $d$.
(c) The correlation between $\Delta V_A$ and $\Delta E_p$.
}
\label{fig12}
\end{figure}

So let us focus on 
the separation between the neighboring CuO$_2$ planes, which will be
denoted as $d$ here.  
Figure \ref{fig12}(a) plots $\Delta V_A^{({\rm O}_c)}$ against  $d$
for La, Hg, Tl, and Bi cuprates. 
Here we have defined $\Delta V_A^{({\rm O}_c)}$ as the contribution to $\Delta V_A$ coming from
the apical oxygens. These Madelung potentials are calculated by 
placing point charges at atomic positions, as was done in ref.\cite{Maekawa}.  
We have also plotted the total $\Delta V_A$ for the four materials, 
which indicates that $\Delta V_A$ is roughly governed by 
$\Delta V_A ^{({\rm O}_c)}$,  
which in turn is mainly determined by $d$.  
We also plot $\Delta E_p$ against $d$  in Fig.\ref{fig12}(b) 
for the four cuprates. Here again, $\Delta E_p$ is 
obtained  using the model that considers the Cu-$3d$ and O-$2p$ orbitals 
explicitly. From these we can see that both 
$\Delta V_A$ and $\Delta E_p$ are primarily correlated positively 
with the layer separation $d$.  This in turn implies that 
$\Delta E_p$ and $\Delta V_A$  in Fig.\ref{fig12}(c) are positively 
correlated as well.

\subsection{Classification of materials by $\Delta E_d$ and $\Delta E_p$}

\begin{figure}[!h]
\includegraphics[width=7cm]{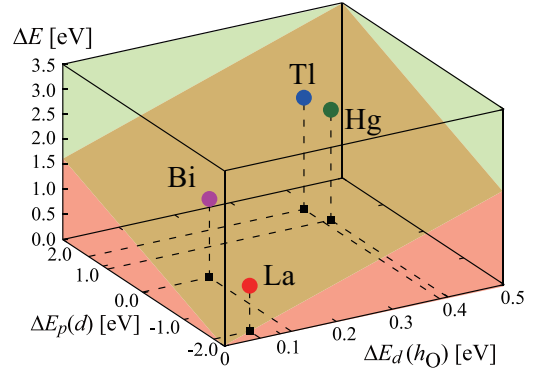}
\caption{(Color online) 
$\Delta E$ plotted against $\Delta E_p$ and $\Delta E_d$ 
for the four single-layered cuprates considered here.  
An oblique plane indicates a 
rough correlation between $\Delta E$ and 
($\Delta E_p, \Delta E_d$).
}
\label{fig13}
\end{figure}

We have seen that 
$\Delta E_d$ and $\Delta E_p$ are mainly determined by $h_{\rm O}$ and
$d$, respectively. Combining these, we can summarize the 
dependence of $\Delta E$ on the material  and lattice structure as
\begin{equation}
\Delta E\simeq f(\Delta E_d(h_{\rm O}),\Delta E_p(d)),
\end{equation}
where $f$ is a certain function.  For instance, 
La and Bi have smaller $\Delta E_d$ reflecting smaller $h_{\rm O}$, while 
Hg and Tl have larger $\Delta E_d$ due to larger $h_{\rm O}$.
Namely, the latter group tends to have larger $\Delta E$.
On the other hand, Bi, Tl, and Hg have larger $d$ than La, so that 
they have larger $\Delta V_A$.
We can summarize all these into a 
classification of materials in terms of $\Delta E_d$ and $\Delta E_p$ 
as a numerical table\ref{table3}, and a kind of ``phase diagram''  
in Fig.\ref{fig13}.  
Apart from the effect of $h_{\rm O}$ (or $\Delta E_d$),  
$\Delta E$ is positively correlated with $\Delta E_p$ and thus with 
$\Delta V_A$, 
so that $\Delta V_A$ and $T_c$ should be roughly correlated.  
In this sense, the so-called Maekawa's plot 
(Fig.2 of ref.\cite{Maekawa}) 
is consistent with the present  Fig.\ref{fig6}. 
Also, a negative correlation between the occupancy of holes 
with $p_z-d_{z^2}$ character and $T_c$ has been found 
in ref.\cite{DiCastro}, which is again consistent with the 
present view.

\begin{table}[!h]
\caption{The values of $\Delta E_d$ (along with $h_{\rm O}$), 
$\Delta E_p$  (along with $d$), and $\Delta E$ for La, Bi, 
Hg and Tl cuprates. }
\begin{tabular}{c c c c c}
\hline
- & \hspace{0.5cm} La \hspace{0.5cm} &\hspace{0.5cm} Bi \hspace{0.5cm} & \hspace{0.5cm} Hg \hspace{0.5cm} & \hspace{0.5cm} Tl \hspace{0.5cm} \\\hline
$\Delta E_d$ [eV] & 0.064 & 0.12  & 0.39 & 0.39 \\
\hspace{0.1cm}$h_{\rm O}\hspace{0.2cm} [{\rm \AA}]$ & 2.41 & 2.46  & 2.78 & 2.71\\\hline
$\Delta E_p$ [eV] & -1.7 & 0.030  & 0.89 & 1.4\\
\hspace{0.2cm}$d$\hspace{0.3cm}[{\rm \AA}] & 6.6 & 12.3  & 9.5 & 11.6 \\
\hline
\hspace{0.05cm}$\Delta E$\hspace{0.05cm} [eV] & 0.91 & 1.6 & 2.2 & 2.2 \\ \hline
\end{tabular}
\label{table3}
\end{table}

\section{Discussions}

\subsection{Validity of the present model}
In the present study, we have adopted 
the LDA  to derive the kinetic-energy part of the model
Hamiltonian. The LDA  calculation neglects some of the 
electron correlation effects, and our standpoint in the present 
study is that the remaining part of the electron correlation 
is dealt with in the FLEX calculation.
One might suspect, however, that there might remain electron 
correlation effects, that are not taken into account in the 
present approach but can affect the accuracy of the evaluation 
of the level offset $\Delta E$ between $d_{x^2-y^2}$ and $d_{z^2}$
Wannier orbitals.  
Our view on this point is the following. 
First, it is an experimental fact that the La cuprate has 
a square-like  Fermi surface, while the Bi cuprate a rounded 
one\cite{Tanaka}. This is accurately reproduced in the 
LDA, which strongly suggests that the $d_{z^2}$ component 
is indeed strongly mixed around $(\pi,0),(0,\pi)$, 
i.e., $\Delta E$ is small, in the La cuprate.  
Secondly, a detailed 
quantitative difference in $\Delta E$ will not affect the 
present conclusions.  To see this, we have performed an LDA+U 
calculation to obtain the kinetic-energy part of the Hamiltonian, 
varying $U$ from 0 to $6$ eV. For La,  the 
considerable $d_{z^2}$-character around $(\pi,0),(0,\pi)$ persists 
even at $U=6$, and the band that intersects the Fermi level is 
only slightly changed, although $\Delta E$  
somewhat increases with $U$.  On the other hand, for Hg $\Delta E$ is 
greatly enhanced by $U$, but this does not significantly affect the 
$d_{x^2-y^2}$ main band, since the $d_{z^2}$ character is 
already absent at $U=0$. 
Applying FLEX to these LDA+U models will result in a double counting of 
the electron correlation effects because FLEX takes account of the 
first-order terms, but 
if we took, for the sake of comparison, the obtained $\Delta E$,  
we would find that a considerable difference in $\lambda$ between 
La ($\lambda\simeq 0.5$) and Hg ($\simeq 0.8$) 
is still present even if we adopt the modified values of $\Delta E$.

\subsection{Possibility of higher-$T_c$ materials}

A consequence of our study is that 
superconductivity in the single-layered cuprates 
is optimized when the system has a single-band nature.  
In such a case (as in Hg cuprate), 
the Fermi surface is rounded due to the effect of the 
Cu 4s orbital.  As mentioned in section\ref{IIA}, the square shaped 
Fermi surface would be more favorable for superconductivity 
for single-orbital systems.  For this very reason 
even the HgBa$_2$CuO$_{4+\delta}$ 
is not fully optimized as a single-layered material.  
Indeed, the hypothetical La cuprate having a 
large $\Delta E$ but with a Fermi surface similar to 
that in the original 
La gives a larger $\lambda$ in the Eliashberg equation as 
we have seen in section \ref{IIID}.  
So we have a bit of a dilemma, since it would be difficult to 
get rid of the effect of the Cu 4s orbital {\it as far as the cuprates 
are concerned}.  Conversely, however, 
we can seek for other materials 
in which the $4s$ orbital is not effective. An example is a  
single-band system consisting of $d_{xy}$ orbitals,  
where the hybridization 
between $d_{xy}$ and $4s$ orbitals is forbidden by symmetry.
In fact, a possible way of realizing a single-band $d_{xy}$ system 
has been proposed in ref.\cite{Arita}.  
Provided that such a system has the band width and the 
electron-electron interaction strength similar to those in the cuprates 
(since too strong or too weak a correlation will degrade superconductivity), 
it can possibly give even higher $T_c$.

\section{Conclusion}

To summarize, we have studied a two-orbital model that 
considers both $d_{x^2-y^2}$ and $d_{z^2}$ Wannier orbitals 
in order to pinpoint the key factors governing 
the material dependence of $T_c$ 
within the single-layered cuprates. We conclude that the $d_{z^2}$ 
orbital mixture on the Fermi surface is significantly degrades superconductivity.  Since the energy difference $\Delta E$ between 
the $d_{x^2-y2}$ and the $d_{z^2}$ governs the mixture as well as 
the shape of the Fermi surface, we identify $\Delta E$ as 
the key parameter in the material dependence of $T_c$ in the 
cuprates.    Since the mixing effect supersede the effect of 
the Fermi surface nesting, 
a small $\Delta E$ results in a suppression of $T_c$ 
despite a square shaped Fermi surface.  
$\Delta E$ is then shown to be 
determined by the energy difference $\Delta E_d$ between the two Cu3d orbitals, and the energy difference $\Delta E_p$ 
between the O$_{\rm plane}$ $p_{\sigma}$ and O$_{\rm apical}$ $p_{z}$, 
both of which are affected by the lattice structure.  
$\Delta E_d$ is a crystal field splitting, which is mainly determined by the 
apical oxygen height, while $\Delta E_p$ 
is found to be primarily governed by the interlayer separation $d$.  
The materials that have highest $T_c$s within the single-layered cuprates, 
Hg are Tl systems, indeed have $\Delta E$ large enough to 
make them essentially single-band.  
On the other hand, there is still room for improvement if we can 
suppress the effect of the Cu $4s$ mixing that 
makes the Fermi surface rounded, which may be realized 
in non-cuprate materials with $U$ similar to the cuprates in magnitude.

\section{ACKNOWLEDGMENTS}

We are grateful to O. K. Andersen and D. J. Scalapino for fruitful discussions.
The numerical calculations were performed at the Supercomputer Center, 
ISSP, University of Tokyo. This study has been supported by 
Grants-in-Aid for Scientific Research from MEXT(Grant No.23340095:R.A.)
 of Japan and from JSPS(Grant No.23009446:H.S., No.21008306:H.U., and No.22340093:K.K. and H.A.). 
H.S. and H.U. acknowledge support from JSPS.
R.A. thanks financial support from JST-PRESTO.


\begin{thebibliography}{99}
\bibitem{comment3} The conclusion of the present study is not 
sensitive to the choice of the interaction values. For instance, 
if we raise the interactions to $U=4.5$ eV $(\sim 10t )$ and $J=0.45$ eV,  
the eigenvalue of the Eliashberg equation 
for the two-orbital model discussed in Fig.
\ref{fig6}
 is only slightly modified to $\lambda=0.29$ for La and $0.72$ for Hg.
\end{thebibliography}
\end{document}